\documentclass[12pt]{l4dc2022_MODDED}

\makeatletter
\def\set@curr@file#1{\def\@curr@file{#1}} 
\makeatother
\usepackage[load-configurations=version-1]{siunitx} 

\title[Robust Data-Driven Output Feedback Control]{Robust Data-Driven Output Feedback Control\\ via Bootstrapped Multiplicative Noise}
\usepackage{times}
\usepackage{enumitem}
\usepackage{algorithm}
\usepackage{algorithmic}

\newcommand{\EE}{\mathbb{E}}
\newcommand{\RR}{\mathbb{R}}
\newcommand{\bbS}{\mathbb{S}}

\newcommand{\tp}{\intercal}

\DeclareMathOperator{\vect}{vec}
\DeclareMathOperator{\mat}{mat}
\DeclareMathOperator{\svec}{svec}
\DeclareMathOperator{\smat}{smat}
\DeclareMathOperator{\Tr}{Tr}

\newcommand{\kron}{\otimes}

\author{%
 \Name{Benjamin Gravell$^*$} \Email{benjamin.gravell@utdallas.edu}\\
 \Name{Iman Shames$^\dag$} \Email{iman.shames@anu.edu.au}\\
 \Name{Tyler Summers$^*$} \Email{tyler.summers@utdallas.edu}\\
 \addr $^*$The University of Texas at Dallas \quad\quad $^\dag$The Australian National University%
}

\begin{document}

\maketitle

\begin{abstract}%
We propose a robust data-driven output feedback control algorithm that explicitly incorporates inherent finite-sample model estimate uncertainties into the control design. The algorithm has three components: (1) a subspace identification nominal model estimator; (2) a bootstrap resampling method that quantifies non-asymptotic variance of the nominal model estimate; and (3) a non-conventional robust control design method comprising a \emph{coupled} optimal dynamic output feedback filter and controller with multiplicative noise. A key advantage of the proposed approach is that the system identification and robust control design procedures both use stochastic uncertainty representations, so that the actual inherent statistical estimation uncertainty directly aligns with the uncertainty the robust controller is being designed against. Moreover, the control design method accommodates a highly structured uncertainty representation that can capture uncertainty \emph{shape} more effectively than existing approaches. We show through numerical experiments that the proposed robust data-driven output feedback controller can significantly outperform a certainty equivalent controller on various measures of sample complexity and stability robustness.%
\end{abstract}


\section{Introduction}
The intersection of data-driven learning and model-based control continues to provide significant research challenges despite its long history and vast research literature. Recent work has focused on non-asymptotic analysis of sample complexity, regret, and robustness, in contrast to a classical focus on asymptotics and stability. Approaches for data-driven control can be broadly divided into two categories: ``model-based'' (or ``indirect''), in which a model for the system dynamics is first learned from data and then used to design a control policy, and ``model-free'' (or ``direct''), in which a control policy is learned directly from data without explicitly learning a model for the system dynamics. Model-based approaches can be further divided into two categories: certainty equivalent, in which uncertainty in the learned model is ignored during control design, and robust, in which uncertainty in the learned model is explicitly accounted for in control design.

Much recent work has considered the full state feedback setting, and some results have very recently been obtained in the partially observed output feedback setting. Finite-sample bounds for system identification from input-state data have been obtained in \cite{simchowitz2018learning, dean2020sample} and from input-output data in \cite{care2017finite,tsiamis2019finite,sun2020finite,jedra2019sample,oymak2021revisiting,sarkar2021finite}. Sample complexity and regret bounds for the Linear Quadratic Gaussian problem are described in \cite{zheng2021sample,zhang2021sample} and \cite{lale2020logarithmic,lale2021adaptive,simchowitz2020improper}, respectively.


In the partially observed setting, issues around robustness to model uncertainty are much more pronounced than in the full state feedback setting, a fact long known in control theory \cite{doyle1978guaranteed}. Certainty equivalent approaches that ignore model uncertainty can lead to fragile designs, while existing approaches the incorporate model uncertainty often utilize very coarse uncertainty representations (e.g., spectral norm balls), even when obtaining order optimal statistical sample complexity or regret rates. A good balance between performance and robustness in practice requires carefully constructed and structured uncertainty representations; just as much effort should go into estimating from data the \emph{shape} (not just size) of model uncertainty as the nominal model itself. 
This becomes especially important as the uncertainty dimension increases: structured uncertainties may have far less volume (in model space) than unstructured ones, thereby enabling superior performance. Developing algorithms with good non-asymptotic performance and robustness properties remains a significant challenge, both in theory and in practice. 
To address this challenge, \cite{gravell20a} proposed a data-driven robust control scheme via bootstrapped multiplicative noise for systems with perfect full state measurements; the present work extends these ideas to the partially observed output feedback setting.


\textbf{Contributions.}
The contributions of the present work are as follows:
\begin{enumerate}[topsep=0pt,itemsep=-1ex,partopsep=1ex,parsep=1ex]
    \item We propose a robust data-driven output feedback control algorithm where the model uncertainty description and robust control design method both use highly structured stochastic uncertainty representations. 
    \item We present a novel semi-parametric bootstrap algorithm for quantifying structured parametric uncertainty in state space models obtained from subspace identification algorithms using input-output data.
    \item We show via numerical experiments that the proposed robust data-driven output feedback controller can significantly outperform a certainty equivalent controller on various measures of sample complexity and stability robustness. 
    We make open-source code implementing the algorithms and experiments freely available.
\end{enumerate}
The algorithm has three components: (1) a subspace identification nominal model estimator; (2) a novel semi-parametric bootstrap resampling method that quantifies non-asymptotic variance of the nominal model estimate; and (3) a non-conventional robust control design method using an optimal linear quadratic coupled estimator-controller with multiplicative noise. This approach provides a natural interface between several highly effective methods from system identification, statistics, and optimal control theory 
(namely, subspace identification, bootstrap resampling, and robust control).

\subsection{Notation}
\begin{center}
  \begin{tabular}{p{2cm} p{12.5cm}} 
  Symbol & Meaning  \\
  \hline
    {$\RR^{n \times m}$} & {Space of real-valued $n \times m$ matrices} \\
    {$\bbS^{n}$} & {Space of symmetric real-valued $n \times n$ matrices} \\
    {$\bbS^{n}_{+}$} & {Space of symmetric real-valued positive semidefinite $n \times n$ matrices} \\
    {$\bbS^{n}_{++}$} & {Space of symmetric real-valued strictly positive definite $n \times n$ matrices} \\
    {$\rho(M)$} & {Spectral radius (greatest magnitude of an eigenvalue) of a square matrix $M$} \\
    {$\lVert M \rVert$} & {Spectral norm (greatest singular value) of a matrix $M$} \\
    {$\lVert M \rVert_F$} & {Frobenius norm (Euclidean norm of the vector of singular values) of a matrix $M$} \\
    {$M \otimes N$} & {Kronecker product of matrices $M$ and $N$} \\
    {$\vect(M)$} & {Vectorization of matrix $M$ by stacking its columns} \\
    {$\mat(v)$} & {Matricization of vector $v$ such that $\mat(\vect(M))=M$} \\
    {$\svec(M)$} & {Symmetric vectorization of matrix $M$ by stacking columns of the upper triangular part, including the main diagonal, with off-diagonal entries multiplied by $\sqrt{2}$ such that $\lVert M \rVert_F^2 = \svec(M)^\tp \svec(M)$} \\
    {$\smat(v)$} & {Symmetric matricization of vector $v$ i.e. inverse operation of $\svec(\cdot)$ such that $\smat(\svec(M))=M$} \\
    {$M \succ (\succeq) \ 0$} & {Matrix $M$ is positive (semi)definite} \\
    {$M \succ (\succeq) \ N$} & {Matrix $M$ succeeds matrix $N$ as $M-N \succ (\succeq) \ 0$ } \\
  \end{tabular}
\end{center}
Let $GL(n)$ denote the general linear group of size $n$, that is the set of non-singular $n \times n$ matrices together with the operation of matrix multiplication.
An ordered sequence of vectors is denoted in the compact notation $x_{0:T} = [x_0, x_1, \ldots, x_T]$.

\section{Problem Formulation: Data-Driven Output Feedback Control}
We consider data-driven control of the discrete-time linear dynamical system
\vspace{-0.1\baselineskip}
\begin{align}
    x_{t+1} &= A x_t + B u_t + w_t,  \vspace{-0.1\baselineskip} \label{eq:true_system1}\\
    y_t     &= C x_t + v_t  \label{eq:true_system2}
\end{align}
where $x_t \in \RR^n$ is the system state, $u_t \in \RR^m$ is the control input, $y_t \in \RR^p$ is the measured output, and $w_t$ and $v_t$ are i.i.d. process and measurement noises with zero mean and covariance matrices $W$ and $V$, respectively. The system matrices $(A,B,C)$ and noise covariances $(W,V)$ are grouped into the true model $\mathcal{M} = (A, B, C, W, V)$ which is assumed unknown.\footnote{We assume the order $n$ of the underlying system is known; future work will address systems with unknown order.} 
Given only on a single training trajectory of finite length $T$ of input-output data $\mathcal{D}_T = (y_{0:T}^{\text{train}}, u_{0:T-1}^{\text{train}})$ generated by the true system \eqref{eq:true_system1}, \eqref{eq:true_system2}, a data-driven input-output history-dependent control policy $u_t = \pi(y_{0:t}, u_{0:t-1})$ is to be designed.
We assume that the input signal that produced the training trajectory was persistently exciting to avoid identifiability issues (see Definition 5 of \cite{van2012subspace}).

The performance of an arbitrary policy $\pi$ is characterized by the infinite-horizon time-averaged linear-quadratic output-input criterion
\begin{align}
    H(\pi) := \lim_{\mathcal{T} \to \infty} \frac{1}{\mathcal{T}} \EE \left[ \sum_{t=0}^{\mathcal{T}-1}  y_t^\tp Y y_t + u_t^\tp R u_t \right] \label{eq:performance_output}
\end{align}
where $Y \succ 0$ and $R \succ 0$ are penalty matrices, $u_t = \pi(y_{0:t}, u_{0:t-1})$, the initial state $x_0$ is a random vector with zero mean and identity covariance independent of the noises $w_t$ and $v_t$, and the expectation is taken with respect to the process and measurement noise sequences and the initial state.
Notice that this formulation permits one to choose the output penalty $Y \succ 0$, which can be specified even if the true underlying state $x_t$ and system model are unknown. The output-input performance criterion \eqref{eq:performance_output} is equivalent, up to a shift by a problem-dependent constant, to a state-input performance criterion with a penalty matrix $Q = C^\tp Y C \succeq 0$, such that
\begin{align}
    J(\pi) := \lim_{\mathcal{T} \to \infty} \frac{1}{\mathcal{T}} \EE \left[ \sum_{t=0}^{\mathcal{T}-1}  x_t^\tp Q x_t + u_t^\tp R u_t \right] = H(\pi) - \Tr(Y V) \label{eq:performance}
\end{align}
so that minimization of $H$ is tantamount to minimization of $J$, which is shifted by a positive constant $\Tr(Y V)$ that does not depend on the policy.

We focus on a sequential design pipeline, in which the data is first used to identify a system model $\hat{\mathcal{M}}(\mathcal{D}_T)$ and then an output feedback control policy $\pi_{\hat{\mathcal{M}}(\mathcal{D}_T)}$ is designed based on the identified model; note that the identified model $\hat{\mathcal{M}}(\mathcal{D}_T)$ is more generic and may have alternative or additional structure compared to the true model $\mathcal{M}$.
A \emph{linear dynamic compensator} is a policy which combines a linear state estimator with a linear state estimate feedback in the form
\begin{align}
    \hat{x}_{t+1} = F \hat{x}_t + L y_t, \quad u_t = K \hat{x}_t. \label{eq:linear_dynamic_compensator}
\end{align}
Such a compensator is fully specified by the triple $(F, K, L)$, and the specification need not depend on the state $x_t$ or system matrices $(A, B, C)$ of the underlying system.
The optimal cost is the constant $J^* = \min_\pi J(\pi) = J(\pi_\mathcal{M})$, which is achieved when the true model $\mathcal{M}$ is known and used in the canonical linear quadratic Gaussian (LQG) control policy $\pi_\mathcal{M}$, a linear dynamic compensator with $F = A + BK - LC$ and gain matrices $(K, L)$ computed (separately) as the solution to two decoupled algebraic Riccati equations, which can be accomplished via several well-known methods such as the dynamic programming techniques of policy iteration and value iteration \cite{bertsekas2012dynamic}, convex semidefinite programming \cite{boyd1994linear}, and specialized direct linear algebraic methods \cite{laub1979}.
Therefore, we restrict attention to the class of linear dynamic compensators in \eqref{eq:linear_dynamic_compensator}.
Using a compensator $(F, K, L)$, the closed-loop system dynamics become the autonomous stochastic difference equation
\begin{align}
    \begin{bmatrix}
    x_{t+1} \\
    \hat{x}_{t+1}
    \end{bmatrix}
    =
    \begin{bmatrix}
    A & BK \\
    LC & F
    \end{bmatrix}
    \begin{bmatrix}
    x_{t} \\
    \hat{x}_{t}
    \end{bmatrix}
    + 
    \begin{bmatrix}
    I & 0 \\
    0 & L
    \end{bmatrix}
    \begin{bmatrix}
    w_{t} \\
    v_{t}
    \end{bmatrix}    
\end{align}
Denote the following augmented closed-loop matrices
\begin{align}
    \Phi = 
    \begin{bmatrix}
    A & BK \\
    LC & F
    \end{bmatrix}, \quad
    Q^\prime =
    \begin{bmatrix}
    Q & 0 \\
    0 & K^\tp R K
    \end{bmatrix}, \quad
    W^\prime =
    \begin{bmatrix}
    W & 0 \\
    0 & L V L^\tp
    \end{bmatrix}.   \label{eq:closed_loop_ldc} 
\end{align}
The stability of the closed-loop system is characterized by the spectrum of the matrix $\Phi$, namely if $\rho(\Phi) < 1$ then the closed-loop system is stable in the sense that the covariance of the augmented state $[ x_t \ \hat{x}_t]^\tp$ converges to a finite positive definite matrix as $t \to \infty$. 
With such stability, the steady-state value matrix $P^\prime$ and the steady-state covariance $S^\prime$ of $[ x_t^\tp \ \hat{x}_t^\tp]^\tp$ are found by solving the discrete-time Lyapunov equations
\begin{align}
    P^\prime = \Phi^\tp P^\prime \Phi + Q^\prime, \quad S^\prime = \Phi S^\prime \Phi^\tp + W^\prime. \label{eq:dlyap_aug}
\end{align}
With a slight abuse of notation, the performance criterion \eqref{eq:performance} can be expressed and computed as
\begin{align}
    J(F, K, L) = \Tr(P^\prime W^\prime) = \Tr(S^\prime Q^\prime). \label{eq:performance_linear}
\end{align}
Denote the performance criterion and closed-loop system matrix under a linear dynamic compensator $(\hat F_T, \hat K_T, \hat L_T)$ designed with the $T$-step data record $\mathcal{D}_T$ as $J_T = J (\hat F_T, \hat K_T, \hat L_T)$ from \eqref{eq:performance_linear} and $\Phi_T$ from \eqref{eq:closed_loop_ldc}, respectively.
The quantity of primary interest is $\frac{J_T}{J^*} \in [1, \infty)$,
which represents the normalized infinite-horizon performance at time $T$.
Since the policy is computed based on a model identified from noisy finite data, the ratio $\frac{J_T}{J^*}$ is a random variable. We are interested in its finite sample behavior and finiteness (which relates to stability robustness); in particular, we would like to know not only in how the mean or median scale with the data length $T$, but also how the upper tails scale. These properties depend on whether and how inherent uncertainty in the identified model is accounted for in the controller design. Certainty equivalent approaches ignore the model uncertainty altogether, which may lead to serious finite sample robustness issues. Here we aim to explicitly incorporate the model uncertainty in the controller design. In particular, we propose a robust data-driven output feedback control algorithm that explicitly accounts for finite-sample model uncertainty in an identified model using a multiplicative noise framework, estimated via the bootstrap.

\section{Robust Control via Bootstrapped Multiplicative Noise}
Our robust data-driven control algorithm is summarized in Algorithm \ref{algorithm:algo_whole}, 
The algorithm has three main components: (1) a subspace identification nominal model estimator; (2) a bootstrap resampling method that quantifies non-asymptotic variance of the nominal model estimate; and (3) a non-conventional robust control design method using an optimal LQG with multiplicative noise.
\begin{algorithm}
\caption{Robust Data-Driven Output Feedback Control}
\begin{algorithmic}[1]
\label{algorithm:algo_whole}
    \REQUIRE single trajectory data $\mathcal{D}_T = (y_{0:T}^{\text{train}}, u_{0:T-1}^{\text{train}})$, number of bootstrap resamples $N_b$, model uncertainty scaling parameter $\gamma$, penalty matrices $Y \succ 0, R \succ 0$
    \small 
	\STATE $(\hat A_T, \hat B_T, \hat C_T, \hat W_T, \hat V_T, \hat U_T, \hat w_{0:T}, \hat v_{0:T}) = \tt{SubspaceID}$ $(y_{0:T}, u_{0:T-1})$
	\STATE $(\hat \Sigma_{A_T}, \hat \Sigma_{B_T}, \hat \Sigma_{C_T}) = \tt{BootstrapModelCovariance}$ $(y_{0:T}, u_{0:T-1}, \hat A_T, \hat B_T, \hat C_T, \hat w_{0:T}, \hat v_{0:T}, N_b)$
	\STATE $(\hat F_T, \hat K_T, \hat L_T) = \tt{MultiNoiseLQG}$$(\hat A_T, \hat B_T, \hat C_T, \hat W_T, \hat V_T, \hat U_T, \hat C_T^\tp Y \hat C_T, R, \hat \Sigma_{A_T}, \hat \Sigma_{B_T}, \hat \Sigma_{C_T}, \gamma)$
\end{algorithmic}
\end{algorithm}

\subsection{Subspace Identification for Nominal Model Estimation}
The first component of the algorithm is a subspace identification algorithm to estimate the unknown system matrices from input-output trajectory data. 
Subspace identification algorithms have been developed and studied for several decades \cite{van2012subspace}.
There are several variations, which all involve constructing block Hankel matrices from the data and estimating certain subspaces via singular value decompositions, from which the system matrices and noise covariances can be retrieved.
Any of these can be used within the proposed framework, but for concreteness we use the so-called N4SID algorithm \cite{van1994}.
Based on the input-output data $(y_{0:T}^{\text{train}}, u_{0:T-1}^{\text{train}})$, the subspace identification algorithm produces a nominal estimate of the system state space matrices and the process and measurement noise covariances: 
\begin{equation}
    (\hat A_T, \hat B_T, \hat C_T,\hat W_T, \hat V_T, \hat U_T) = {\tt{SubspaceID}} (y_{0:T}^{\text{train}}, u_{0:T-1}^{\text{train}}).
\end{equation}
Due to space constraints we refer readers to the literature, e.g. ``Combined Algorithm 1'' in Chapter 4 of \cite{van2012subspace}, for details of the subspace identification algorithm.

Due to non-uniqueness of state space representations, the system matrices are estimated within a similarity transformation of an underlying unknown representation. 
Based on the input-output data and the estimated system matrices, subspace algorithms generate \emph{residuals} of the process and measurement noises
$\{ \hat w_\tau \}_{\tau = 0}^{t-1}$, $\{ \hat v_\tau \}_{\tau = 0}^{T-1}$
from which sample average covariance estimates
$
\begin{bmatrix}
\hat W_T & \hat U_T \\
\hat U_T^\tp & \hat V_T
\end{bmatrix}    
$
are produced. Because the estimated system matrices $(\hat A_T, \hat B_T, \hat C_T)$ do not share a state coordinate system with the true system matrices $(A, B, C)$, even though the true cross-covariance between $w_t$ and $v_t$ is assumed zero, the cross-covariance of the estimates disturbances $\hat{w}_t$ and $\hat{v}_t$ may be non-zero and must be estimated and accounted for in the compensator design.

\subsection{Bootstrap Resampling to Quantify Non-Asymptotic Model Uncertainty}
There are inevitably errors in model estimates obtained from subspace identification using any finite data record, due to the process and measurement noises. It is difficult to analytically characterize non-asymptotic uncertainty in these estimates.
Quantifying uncertainty in subspace identification estimates has been considered in \cite{viberg1991statistical,bauer1999consistency,bauer2000analysis,reynders2008uncertainty}, which focus on asymptotic results. 
Bootstrapping has been used to quantify non-asymptotic uncertainty in \cite{bittanti2000bootstrap} for input-output quantities such as frequency response or pole locations.
However, to our best knowledge, these uncertainty quantifications have not been used for control design.

To quantify non-asymptotic uncertainty in the model estimate, we propose a novel semi-parametric time series bootstrap resampling procedure.
In semi-parametric methods, bootstrap data are simulated from the nominal model with the process and measurement noise sampled i.i.d. with replacement from residuals calculated with the nominal model \cite{hardle2003bootstrap}.
Dependence in the data is preserved by construction. 
There are also purely parametric and non-parametric versions of the bootstrap. 
Generally, the semi- and nonparametric bootstraps are less sensitive to assumptions about the model and the noise distribution, while the semi- and pure parametric bootstraps have better small sample performance when the model is correctly specified.
The bootstrap resamples allow for various estimates of finite-sample uncertainty associated with the nominal model; here, we will utilize an estimate of the covariance of the model parameters.
For concreteness, a semi-parametric bootstrap with resampled residuals discussed above is summarized in Algorithm \ref{algorithm:algo_covariance_estimation}.

\paragraph{State-space Alignment} \ \\
Due to non-uniqueness of state space representations, the uncertainty representation should not be obtained directly from a sample covariance of the bootstrap resamples. Instead, for each resample we first find a similarity transformation that minimizes the total squared error to the nominal state space model, and then compute a sample covariance in the transformed coordinates.
Ideally, we would form and solve the following optimization problem
\begin{align}
    \min_{T \in GL(n)} \ \tilde{d}(T) = \psi_A \| T \bar{A} T^{-1} - \hat{A} \|^2_F + \psi_B \| T \bar{B} - \hat{B} \|^2_F + \psi_C \| \bar{C} T^{-1} - \hat{C} \|^2_F \label{eq:alignment_problem_target}
\end{align}
which attempts to bring the source model $(\bar{A}, \bar{B}, \bar{C})$ as close to the nominal model $(\hat{A}, \hat{B}, \hat{C})$ as possible by selecting the decision matrix $T$ that defines the coordinate transformation. 
Notice that the coordinates of the nominal model $(\hat{A}, \hat{B}, \hat{C})$ are treated as a ground truth to which the source model $(\bar{A}, \bar{B}, \bar{C})$ should be aligned; reversing their roles would yield a different, less meaningful transformation.
The constants $\psi_A$, $\psi_B$, $\psi_C$ are user-selected to tune the relative weight of the alignment of $A$, $B$, and $C$. For simplicity these are set to unity; further tuning of these constants is left to future work.  
This problem has been referred to as the ``realization alignment'' problem; however, as noted by \cite{jimenez2013}, there are several mathematical issues which complicate solving this problem to global optimality, including the non-compactness of $GL(n)$ and nonconvexity of $\tilde{d}$.
The approach developed in \cite{jimenez2013} to address these issues is specialized to a certain class of LTI systems, namely those with $C$ full column rank, which may not include the LTI systems which result from the subspace identification algorithm we use in this work, and is therefore not appropriate for the current setting.

As an alternative, we use a slightly different objective which does not involve the inverse of the transform matrix $T$, and is in fact linear in the transform matrix $T$ and is no longer constrained to $GL(n)$:
\begin{align}
    \min_{T \in \RR^{n \times n}} \ d(T) = \psi_A \| T \bar{A} - \hat{A} T \|^2_F + \psi_B \| T \bar{B} - \hat{B} \|^2_F + \psi_C \| \bar{C} - \hat{C} T \|^2_F \label{eq:alignment_problem_primal}
\end{align}
Another alternative is the dual problem
\begin{align}
    \min_{T^{-1} \in \RR^{n \times n}} \ d_{\text{dual}}(T^{-1}) = \psi_A \| \bar{A} T^{-1} - T^{-1} \hat{A} \|^2_F + \psi_B \| \bar{B} - T^{-1} \hat{B} \|^2_F + \psi_C \| \bar{C} T^{-1} - \hat{C} \|^2_F \label{eq:alignment_problem_dual}
\end{align}
where the inverse transform matrix $T^{-1}$ is used as the decision variable instead.
Notice that the solution of \eqref{eq:alignment_problem_dual} is the same as that of a problem of the form of \eqref{eq:alignment_problem_primal} with the roles of the target $(\hat{A}, \hat{B}, \hat{C})$ and source $(\bar{A}, \bar{B}, \bar{C})$ models reversed. 
However, in general the solutions $T$ to each of the problems \eqref{eq:alignment_problem_target}, \eqref{eq:alignment_problem_primal} and \eqref{eq:alignment_problem_dual} are not the same; the choice between \eqref{eq:alignment_problem_primal} and \eqref{eq:alignment_problem_dual} is somewhat arbitrary, so we choose the former.
The problem \eqref{eq:alignment_problem_primal} is unconstrained, smooth, and strictly convex; as such there is a unique global minimizer located at the stationary point where the derivative of the objective vanishes. Explicitly, the derivative of the objective can be found by expressing the objective in terms of the trace as
\begin{align*}
    d(T) & = \psi_A \Tr \left[ (T \bar{A} - \hat{A} T)^\tp (T \bar{A} - \hat{A} T) \right] 
           + \psi_B \Tr \left[ ( T \bar{B} - \hat{B} )^\tp ( T \bar{B} - \hat{B} ) \right] \\
           & \quad + \psi_C \Tr \left[ (\bar{C} - \hat{C} T)^\tp (\bar{C} - \hat{C} T) \right]
\end{align*}
then using standard matrix derivative rules e.g. \cite{petersen2012cookbook} to obtain the derivative
\begin{align*}
    \frac{\partial d}{\partial T} 
    & = 2 \psi_A \left( T \bar{A} \bar{A}^\tp + \hat{A}^\tp \hat{A} T - \hat{A}^\tp T \bar{A} - \hat{A} T \bar{A}^\tp \right) \\
    & \quad + 2 \psi_B \left( T \bar{B} \bar{B}^\tp - \hat{B} \bar{B}^\tp \right)
      + 2 \psi_C \left( \hat{C}^\tp \hat{C} T - \hat{C}^\tp \bar{C} \right)
\end{align*}
Setting the derivative equal to zero yields a linear matrix equation in $T$, in fact a kind of generalized Lyapunov equation, which can be solved e.g. via vectorization and Kronecker products and solution of a linear vector equation:
\begin{align}
    T = \mat\left[ G^{-1}  \vect(H) \right]
\end{align}
where $G$ and $H$ are the matrices
\begin{align*}
    G &= \psi_A \left( \bar{A} \bar{A}^\tp \kron I_n + I_n \kron \hat{A}^\tp \hat{A} - \bar{A}^\tp \kron \hat{A}^\tp - \bar{A} \kron \hat{A} \right) \\
    & \quad + \psi_B \left( \bar{B} \bar{B}^\tp \kron I_n \right) + \psi_C \left( I_n \kron \hat{C}^\tp \hat{C} \right), \\
    H &= \psi_B \hat{B} \bar{B}^\tp + \psi_C \hat{C}^\tp \bar{C}.
\end{align*}
It is assumed that $G$ is invertible so that this equation is solvable and results in an invertible transformation matrix $T$.
Then the transformed system matrices are computed as
\begin{align}
    \tilde{A} = T \bar{A} T^{-1}, \quad
    \tilde{B} = T \bar{B}, \quad 
    \tilde{C} = \bar{C} T^{-1}, \quad 
    \tilde{W} = T \bar{W} T^\tp, \quad 
    \tilde{V} = \bar{V}, \quad
    \tilde{U} = T \bar{U}.
\end{align}
Note that in the special case when the nominal model $(\hat{A}, \hat{B}, \hat{C})$ and the source model $(\bar{A}, \bar{B}, \bar{C})$ are related exactly by a similarity transformation, the solution $T$ to the optimization problem is precisely this similarity transform, the optimal objective value is identically zero, and we obtain exact matching $(\tilde{A}, \tilde{B}, \tilde{C}) = (\hat{A}, \hat{B}, \hat{C})$.
This coordinate alignment is incorporated into the model covariance estimation Algorithm \ref{algorithm:algo_covariance_estimation}.

\begin{algorithm}
\caption{Semi-parametric Bootstrap Model Covariance Estimation}
\begin{algorithmic}[1]
\label{algorithm:algo_covariance_estimation}
    \REQUIRE trajectory data $(y_{0:t}, u_{0:t-1})$, nominal model estimate $(\hat A_t, \hat B_t, \hat C_t)$, residuals $\{ \hat w_\tau \}_{\tau = 0}^t, \{ \hat v_\tau \}_{\tau = 0}^t$, number of bootstrap resamples $N_b$
    \STATE $\bar x_0 = \hat x_0$
    \STATE $\bar u_{0:t-1} = u_{0:t-1}$ \;    
    \FOR{$k=1,\ldots,N_b$}
        \STATE Generate data $\bar x_{\tau+1} = \hat A_t \bar x_\tau + \hat B_t \bar u_\tau  + \tilde w_\tau$, $\bar y_\tau = \hat C_t \bar x_\tau + \tilde v_{\tau}$, $\tau=0,...,t-1$, where $\tilde w_{0:t-1}$ and $\tilde v_{0:t-1}$ are i.i.d. resamples with replacement from residuals $\hat w_{0:t-1}$ and $\hat v_{0:t-1}$  \;
        \STATE $ (\hat A_t^k, \hat B_t^k, \hat C_t^k, -, -, -, -, -) = \tt{SubspaceID}$$(\bar y_{0:t}, \bar u_{0:t-1})$
        \STATE $T^* = \arg \min_{T\in \RR^{n \times n}}  \| T \hat A_t - \hat A_t^k T \|_F^2 + \| T \hat B_t - \hat B_t^k \|_F^2 + \| \hat C_t - \hat C_t^k T \|_F^2 $ \;
        \STATE $ \tilde A_t^k =  T^* \hat A_t T^{*-1}, \quad \tilde B_t^k = T^* \hat B_t,\quad \tilde C_t^k =  \hat C_t T^{*-1}$
    \ENDFOR
    \ENSURE Bootstrap sample covariance $\hat \Sigma_{A_t} = \frac{1}{N_b-1} \sum_{k=1}^{N_b} \text{vec}(\tilde A_t^k - \hat A_t) \text{vec}(\tilde A_t^k - \hat A_t)^\tp$ \\
    \qquad \ Bootstrap sample covariance $\hat \Sigma_{B_t} = \frac{1}{N_b-1} \sum_{k=1}^{N_b} \text{vec}(\tilde B_t^k - \hat B_t) \text{vec}(\tilde B_t^k - \hat B_t)^\tp$ \\
    \qquad \ Bootstrap sample covariance $\hat \Sigma_{C_t} = \frac{1}{N_b-1} \sum_{k=1}^{N_b} \text{vec}(\tilde C_t^k - \hat C_t) \text{vec}(\tilde C_t^k - \hat C_t)^\tp$  \;
\end{algorithmic}
\end{algorithm}



\subsection{Multiplicative Noise LQG: Combined Controller and State Estimator}
The model covariance estimate generated from bootstrap resampling interfaces quite naturally with a variant of the optimal linear quadratic output feedback controller that incorporates multiplicative noise, which has a long history in control theory but is far less widely known than its additive noise counterpart (\cite{Wonham1967, bernstein1986robust, dekoning1992, gravell2019learning}). Consider the optimal control problem to find an output feedback controller $u_t = \pi(y_{0:t})$ for dynamics perturbed by multiplicative noise
\begin{alignat}{2}  \label{eq:mLQR}
    &\underset{{\pi \in \Pi}}{\text{minimize}} \quad &&   \lim_{\mathcal{T} \rightarrow \infty} \frac{1}{\mathcal{T}} \EE \sum_{t=0}^{\mathcal{T}-1} (x_t^\tp Q x_t + u_t^\tp R u_t), \\
    &\text{subject to}                         \quad &&x_{t+1} = (A  + \bar A_t) x_t +  (B + \bar B_t) u_t + w_t, \\
    & && \quad y_t = (C  + \bar C_t) x_t + v_t \nonumber
\end{alignat}
where $\bar A_t$, $\bar B_t$, and $\bar C_t$ are i.i.d. zero-mean random matrices with a joint covariance structure over their entries governed by the covariance matrices $\Sigma_A := \EE [\vect(\bar A)\vect(\bar A)^\tp ] \in \RR^{n^2 \times n^2}$, $\Sigma_B := \EE [ \vect(\bar B)\vect(\bar B)^\tp ] \in \RR^{nm \times nm}$, $\Sigma_C := \EE [ \vect(\bar C)\vect(\bar C)^\tp ] \in \RR^{pn \times pn}$ which quantify uncertainty in the nominal system matrices $(A,B,C)$. The expectation is taken with respect to all of the basic random quantities in the problem, namely ${x_0,\{\bar A_t\}, \{\bar B_t \}, \{\bar C_t \}, \{ w_t \}, \{ v_t \}}$.

Due to the multiplicative noise, the state distribution is non-Gaussian even when all primitive distributions are Gaussian, so the Kalman filter is not necessarily the optimal state estimator. However, the optimal \emph{linear} output feedback controller can be exactly computed, and consists of a multiplicative noise linear dynamic compensator of the form \eqref{eq:linear_dynamic_compensator}.
In this case, \emph{there is no separation between estimation and control}, so the optimal controller and estimator gains $(K, L)$ must be jointly computed. 
Specifically, the optimal gains can be computed by solving the coupled nonlinear matrix equations in symmetric matrix variables $X = (X_1, X_2, X_3, X_4)$
\begin{align} \label{eq:genriccati} 
    X_1   & = Q + A^\tp X_1 A + \sum_{i=1}^{n^2} \alpha_i A_i^\tp X_1 A_i - K^\tp \left( R + B^\tp X_1 B + \sum_{j=1}^{nm} \beta_j B_j^\tp X_1 B_j + \sum_{j=1}^{nm} \beta_j B_j^\tp X_2 B_j \right) K \nonumber \\
    & \quad + \sum_{i=1}^{n^2} \alpha_i A_i^\tp X_2 A_i + \sum_{i=1}^{pn} \lambda_i C_i^\tp L^\tp X_2 L C_i \nonumber \\
    X_2 & = (A - LC)^\tp X_2 (A - LC) + K^\tp \left( R + B^\tp X_1 B + \sum_{j=1}^{nm} \beta_j B_j^\tp X_1 B_j + \sum_{j=1}^{nm} \beta_j B_j^\tp X_2 B_j \right) K \nonumber \\
    X_3 & = W + A X_3 A^\tp + \sum_{i=1}^{n^2} \alpha_i A_i X_3 A_i^\tp - L \left( V + C X_3 C^\tp + \sum_{j=1}^{pn} \lambda_j C_j X_3 C_j^\tp + \sum_{j=1}^{pn} \lambda_j C_j X_4 C_j^\tp \right) L^\tp \nonumber \\
    & \quad + \sum_{i=1}^{n^2} \alpha_i A_i X_4 A_i^\tp + \sum_{i=1}^{nm} \beta_i B_i K X_4  K^\tp B_i^\tp \nonumber \\
    X_4 & = (A + BK) X_4 (A + BK)^\tp + L \left( V + C X_3 C^\tp + \sum_{j=1}^{pn} \lambda_j C_j X_3 C_j^\tp + \sum_{j=1}^{pn} \lambda_j C_j X_4 C_j^\tp \right) L^\tp
\end{align}
where $\{\alpha_i, A_i\}_{i=1}^{n^2}$, $\{\beta_j, B_j\}_{j=1}^{nm}$, and $\{\lambda_j, C_j\}_{j=1}^{pn}$ are the eigenvalues and reshaped eigenvectors of $\Sigma_A$, $\Sigma_B$, and $\Sigma_C$, respectively, and
\begin{align} \label{eq:genriccati_gains}
    K & = -\left( R + B^\tp X_1 B + \sum_{j=1}^{nm} \beta_j B_j^\tp X_1 B_j + \sum_{j=1}^{nm} \beta_j B_j^\tp X_2 B_j \right)^{-1} B^\tp X_1 A \\
    L & = (U + A X_3 C^\tp) \left( V + C X_3 C^\tp + \sum_{j=1}^{pn} \lambda_j C_j X_3 C_j^\tp + \sum_{j=1}^{pn} \lambda_j C_j X_4 C_j^\tp \right)^{-1}
\end{align}
The associated optimal cost is then given by 
\begin{equation*}
    J^* = \Tr(Q X_3 + (Q + K^\tp R K) X_4) = \Tr (W X_1 + (W + L V L^\tp) X_2 )
\end{equation*}
These equations are solved using a value iteration algorithm, described in \cite{dekoning1992}. 
In the absence of multiplicative noise, they reduce to the familiar separated algebraic Riccati equations for optimal estimation and control.
The solutions are denoted
\begin{equation} \label{gdare}
    (X, K, L) = \texttt{GDARE}(A, B, C, W, V, U, Q, R, \Sigma_{A}, \Sigma_{B}, \Sigma_{C})
\end{equation}
Both the optimal controller and estimator gains depend explicitly on the model uncertainty, as quantified by the variances of the system matrices, as well as the process and measurement noise covariances. 
This policy is known to provide robustness to uncertainties in the parameters of the nominal model (\cite{bernstein1986robust}). 
Furthermore, the uncertainty in the nominal model estimate used in this control design method is richly structured and derived directly from the finite available data.

In the proposed data-driven control algorithm, we simply substitute the estimated nominal model and model covariance matrices obtained from the subspace identification and bootstrap methods into the multiplicative noise compensator design equations. We also introduce a parameter $\gamma$ which provides a fixed scaling of the model uncertainty. Note that $\gamma = 0$ corresponds to certainty equivalent control, and as $\gamma$ increases, more weight is placed on uncertainty in the nominal model. For $\gamma \in (0,1)$, this approach can be interpreted as shrinkage estimation of the model sample covariance matrices towards certainty equivalence \cite{ledoit2004honey}. Existence of a solution to the generalized Riccati equation depends not just on stabilizability and detectability of the nominal system $(A,B,C)$, but also on the \emph{mean-square stabilizability} via dynamic output feedback of the multiplicative noise system (called \emph{mean-square compensatability} in \cite{dekoning1992}). When the multiplicative noise variances are too large, it may be impossible to stabilize the system in the mean-square sense.
In this case, we scale down the model variances to compute a mean-square stabilizing dynamic output feedback controller; see Algorithm \ref{algorithm:algo_mlqg}.
In particular, we verify the system with specified $\gamma$ is mean-square stabilizable by checking whether the generalized Riccati equation admits a positive semidefinite solution; if not, we find the upper limit $\gamma_\text{max} = c_\gamma \gamma$ via bisection (e.g. \cite{burden1978numerical}) on a scaling $c_\gamma \in [0,1]$.

\begin{algorithm}
\caption{Multiplicative Noise LQG}
\begin{algorithmic}[1]
\label{algorithm:algo_mlqg}
    \REQUIRE Nominal model matrices $A$, $B$, $C$, additive disturbance covariances $W$, $V$, $U$, penalty matrices $Q$, $R$, covariances $\Sigma_A, \Sigma_B, \Sigma_C$, scaling $\gamma$, bisection tolerance $\epsilon > 0$
    \STATE Find largest $c_{\gamma} \in [0,1]$ via bisection such that there exists a feasible solution to \eqref{gdare} \\
    \STATE $(X, K, L) = \texttt{GDARE}(A, B, C, W, V, U, Q, R, c_{\gamma} \gamma \Sigma_A, c_{\gamma} \gamma \Sigma_B, c_{\gamma} \gamma \Sigma_C)$ \;
    \ENSURE $(A + BK - LC, K, L)$
\end{algorithmic}
\end{algorithm}

\section{Numerical Experiments}
We examined the following 2-state shift register with system, penalty, and noise covariance matrices
\begin{align*}
    \left[
    \arraycolsep=2pt\def\arraystretch{1.0}
    \begin{array}{c|c}
    A & B \\ \hline
    C & \
    \end{array}\right]
    =
    \left[
    \arraycolsep=2pt\def\arraystretch{1.0}
    \begin{array}{c c|c}
    0 & 1 & 0 \\
    0 & 0 & 1 \\ \hline
    1 & -1 & \
    \end{array}\right], \quad
    \left[
    \arraycolsep=2pt\def\arraystretch{1.0}
    \begin{array}{c|c}
    Q & \ \\ \hline
    \ & R \
    \end{array}\right]
    =
    \left[
    \arraycolsep=2pt\def\arraystretch{1.0}
    \begin{array}{c c|c}
    1 & -1 & \ \\
    -1 & 1 & \ \\ \hline
    \ & \ & 0.01
    \end{array}\right], \quad
    \left[
    \arraycolsep=2pt\def\arraystretch{1.0}
    \begin{array}{c|c}
    W & \ \\ \hline
    \ & V \
    \end{array}\right]
    =
    \left[
    \arraycolsep=2pt\def\arraystretch{1.0}
    \begin{array}{c c|c}
    0.1 & 0 & \ \\
    0 & 0.1 & \ \\ \hline
    \ & \ & 0.1
    \end{array}\right]
\end{align*}
where the output penalty was $Y = 1$ leading to the given value for $Q=C^\tp Y C$.
The first state stores the previous value of the second state, the second state is determined solely by the control input, and the output is the difference of the two states.
This system is based on the one described in \cite{recht2021blog}, wherein it was shown that the system is extremely sensitive to model identification errors. In particular, despite the open-loop system being perfectly stable with zero eigenvalues, the system under optimal linear quadratic state feedback control is nearly unstable such that any small error in the estimated system matrices produce an unstable closed-loop system. Therefore, this system is likely to see a benefit from the proposed robust control synthesis approach.

The training data $\mathcal{D}_T$ were generated by initializing the state at the origin, applying random controls distributed according to a Gaussian distribution with zero-mean and scaled identity covariance where the scaling was equal to the sum of the largest singular values of $W$ and $V$ (to ensure a sufficiently strong signal-to-noise ratio), and simulating the evolution of the state with the additive process and measurement noise specified by the problem data $(W, V)$.

For brevity, we abbreviate the control design schemes ``certainty-equivalent control'' as ``CE'' and ``robust control via multiplicative noise'' as ``RMN''.
To evaluate the performance of RMN relative to CE, we performed Monte Carlo trials to estimate the distribution of several key quantities: infinite-horizon performance, spectral radius of the closed-loop system, model error, and multiplicative noise variances.
In each Monte Carlo trial, the actual additive noise disturbances $w_t, v_t$ were drawn independently.
The level of additive noise was significant enough that an appreciable number of model estimates remained poor for many timesteps, highlighting the behavior of CE and RMN in the critical high-uncertainty regime.
We simulated the system and evaluated quantities for the trajectory lengths $T \in \{20, 40, 80, 160, 320\}$ according to Algorithm \ref{algorithm:algo_whole}; all of the trajectory lengths were sufficiently long to ensure the estimates in subspace identification were non-degenerate.
We drew $N_{s} = 100,000$ independent Monte Carlo samples and $N_{b} = 100$ bootstrap samples at each time step for uncertainty estimation.
We used unity scaling of the multiplicative noise ($\gamma = 1$) and a tolerance of $\epsilon = 0.01$ for bisection to find the largest scaling $c_\gamma$ of multiplicative noise variance in the multiplicative noise LQG algorithm.

\begin{figure}[htbp!]
    \centering
    \includegraphics[width=0.9\linewidth]{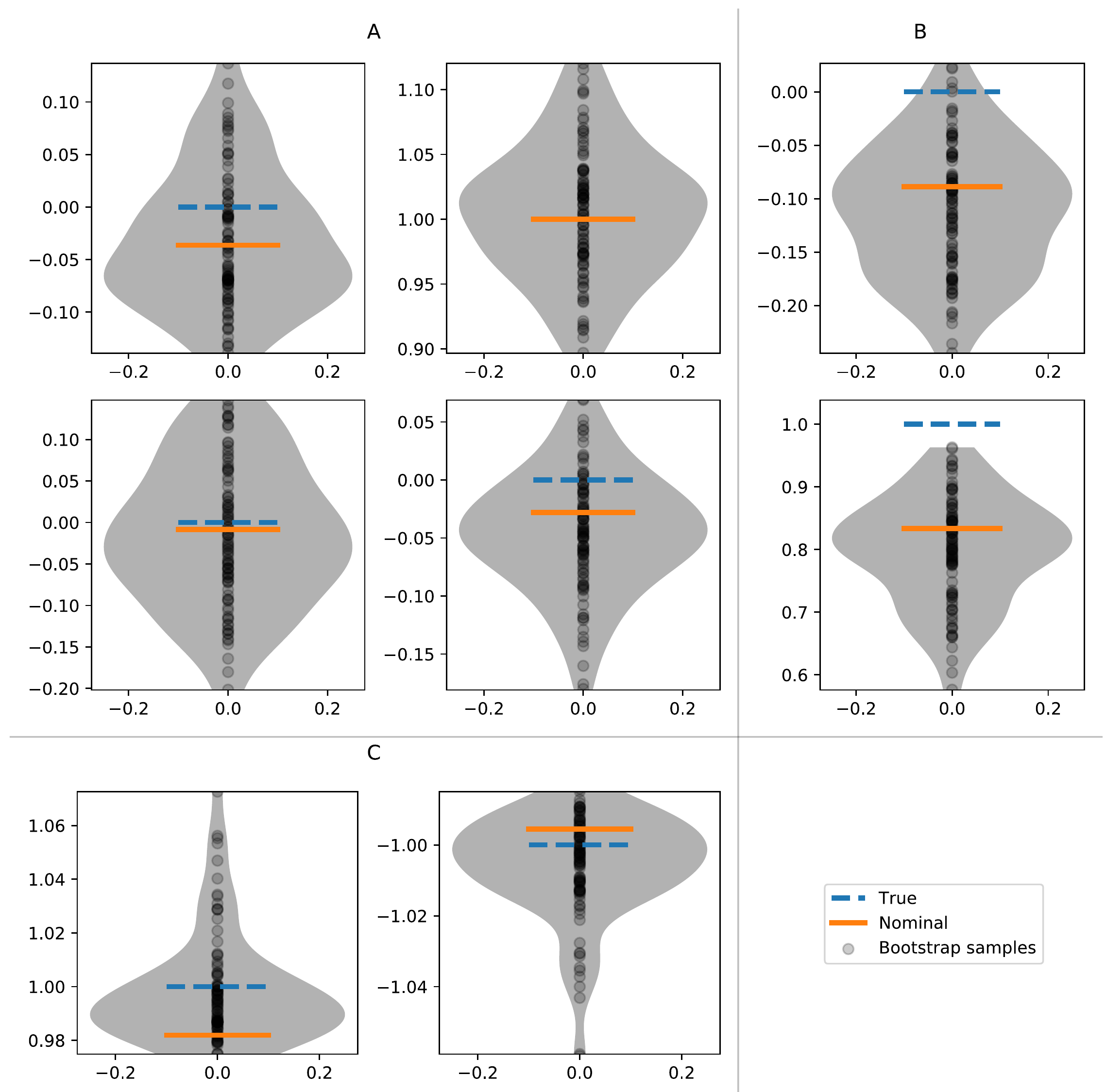}
    \caption{Entrywise plot of the true system, nominal model, and 100 bootstrap samples. Each subplot represents an entry in the system matrices $A$, $B$, and $C$ and are arranged according to the block matrix representation of the system. Entrywise bootstrap distributions estimated using kernel density estimation are shown in shaded regions; these are not used in the proposed approach, they are just shown here for visual clarity. The nominal and bootstrap model matrices were aligned with the true system via a minimum norm transformation; note that this procedure is not part of the proposed approach, and is performed here strictly for comparison with the true system. The data are drawn from the results of just one out of the $N_{s} = 100,000$ Monte Carlo samples for a trajectory length of $T=20$.}
    \label{fig:boot_mat_dist}
\end{figure}

From Figure \ref{fig:boot_mat_dist} we see that for $T=20$ the nominal model is fairly accurate but clearly mis-specified, and that the bootstrap distribution of models captures the true deviation of the nominal model from the true system, as the true system parameters fall within the distribution of bootstrap samples. 
This is an accurate representative of the $N_{s} = 100,000$ Monte Carlo samples.

\begin{figure}[htbp]
    \centering
    \includegraphics[width=0.99\linewidth]{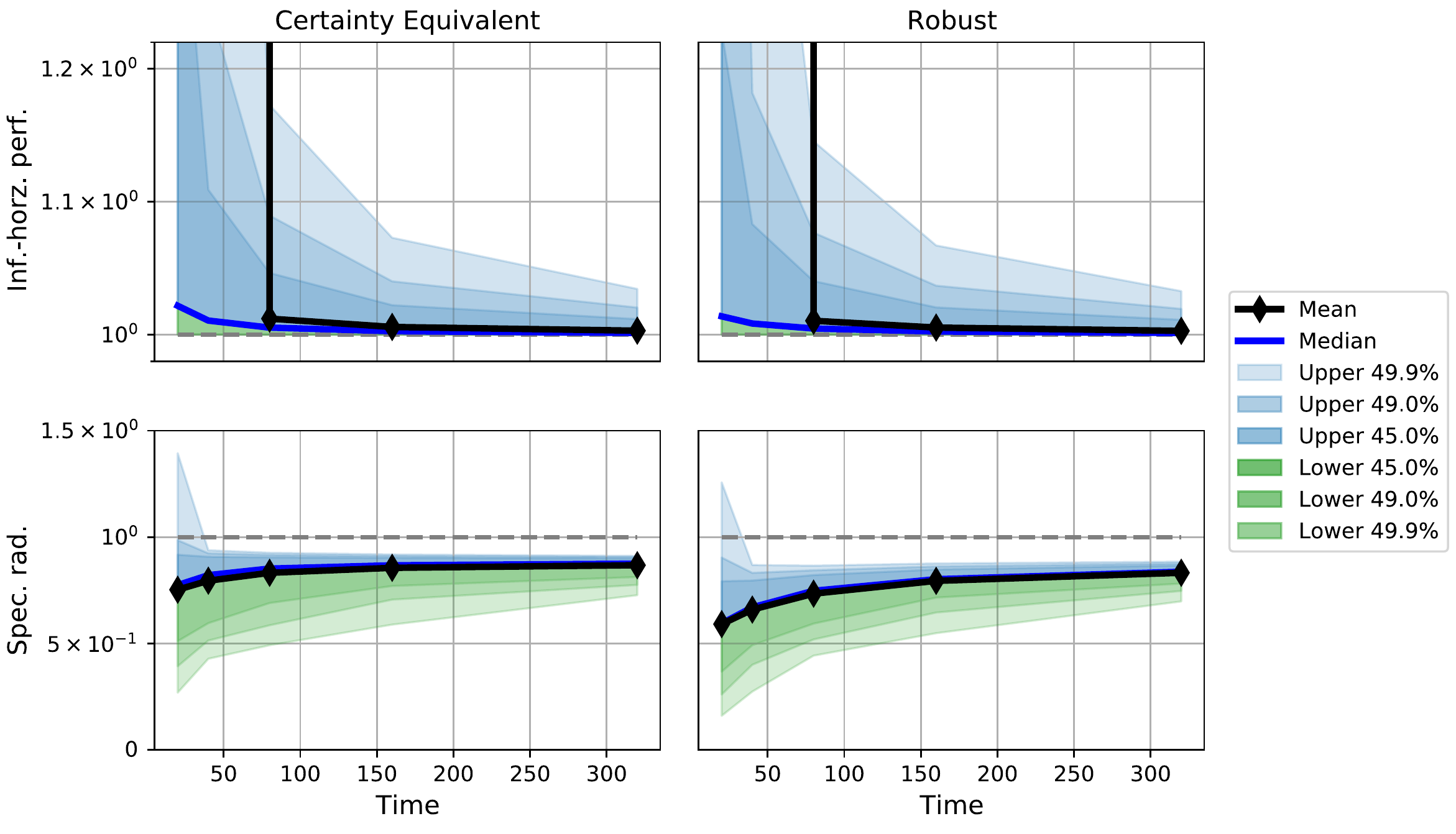}
    \caption{Infinite-horizon performance $J_T / J^*$ and closed-loop spectral radius $\rho(\Phi_T)$ vs time for CE and RMN.}
    \label{fig:perf_and_specrad}
\end{figure}

\begin{figure}[htbp]
    \centering
    \includegraphics[width=0.99\linewidth]{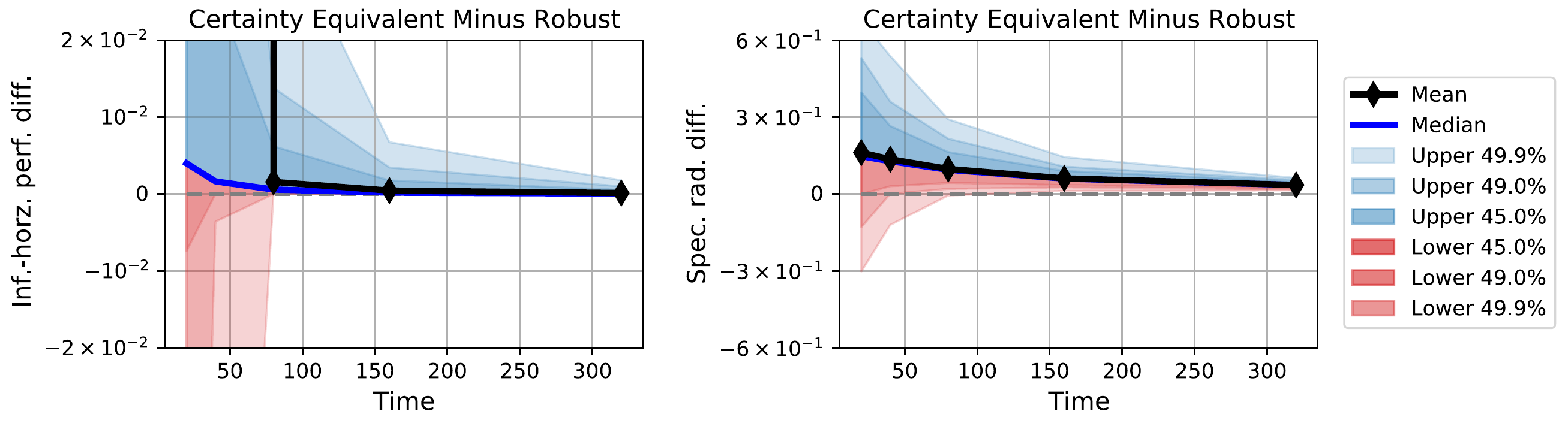}
    \caption{Difference between CE and RMN on infinite-horizon performance $J_T / J^*$ and closed-loop spectral radius $\rho(\Phi_T)$ metrics vs time. Differences between metrics using CE and RMN were computed for each Monte Carlo trial individually; statistics of the resulting empirical distribution are shown.}
    \label{fig:diff_perf_and_specrad}
\end{figure}

In Figure \ref{fig:perf_and_specrad} we plot statistics of performance and spectral radius using both control schemes, while in Figure \ref{fig:diff_perf_and_specrad} we plot statistics of the differences between the performance and spectral radius.
We are chiefly interested in the expected value and upper quantiles of performance, which correspond to average performance and risk of poor performance. We see that CE leads to both worse average behavior and riskier behavior as reflected by the distribution of the performance. In particular, we see that the performance of RMN is clearly better during times between $T=20$ and $T=80$, dropping at the 99th percentile from $2.318$ to $1.077$ whereas CE suffers $12.091$ to $1.089$.
This can also be explained from the spectral radius, which is larger across all timesteps and statistics, corresponding to a less stable system. In particular, at the beginning between $T=20$ and $T=80$ when uncertainty is highest, RMN yields spectral radii at the 99th percentile that drop from $0.903$ to $0.843$ while CE yields $0.985$ to $0.916$, nearer to instability and allowing the state to travel far from the origin, resulting in high cost.
With increasing $T$ the model estimates improved and uncertainty estimates became sufficiently small that the difference between CE and RMN control was insignificant.

\begin{figure}[htbp]
    \centering
    \includegraphics[width=0.99\linewidth]{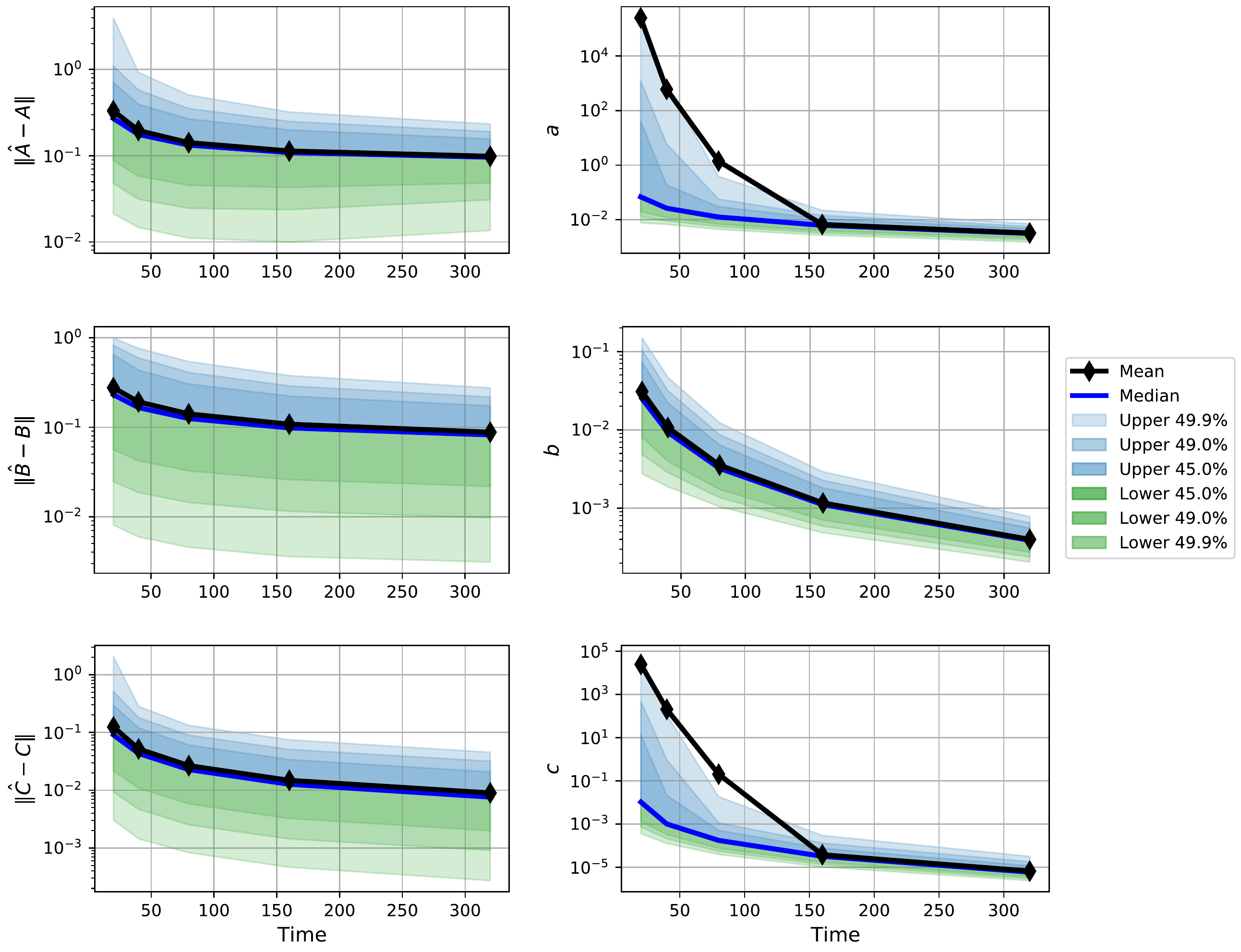}
    \caption{System matrix estimation errors and multiplicative noise variances vs time using RMN. The left column shows the Frobenius norm of the estimated system matrices from the true parameters after applying an alignment transformation. The right column shows the maximum multiplicative noise variances $a = \max_i \alpha_i$, $b = \max_i \beta_i$, $c = \max_i \lambda_i$ at each time step.}
    \label{fig:matrix_errors_and_noises}
\end{figure}

In Figure \ref{fig:matrix_errors_and_noises} we plot statistics of the nominal model estimate errors, which are applicable to both control schemes. We see that the nominal system matrices $\hat{A}$, $\hat{B}$, and $\hat{C}$ produced by the subspace identification algorithm approached the true parameters (after a suitable alignment transformation). This is mirrored by the decrease in the multiplicative noise variances, showing that the multiplicative noise variances accurately reflect the true model error, i.e., the bootstrap model uncertainty estimator gives reasonable estimates.

\begin{figure}[htbp]
    \centering
    \includegraphics[width=0.4\linewidth]{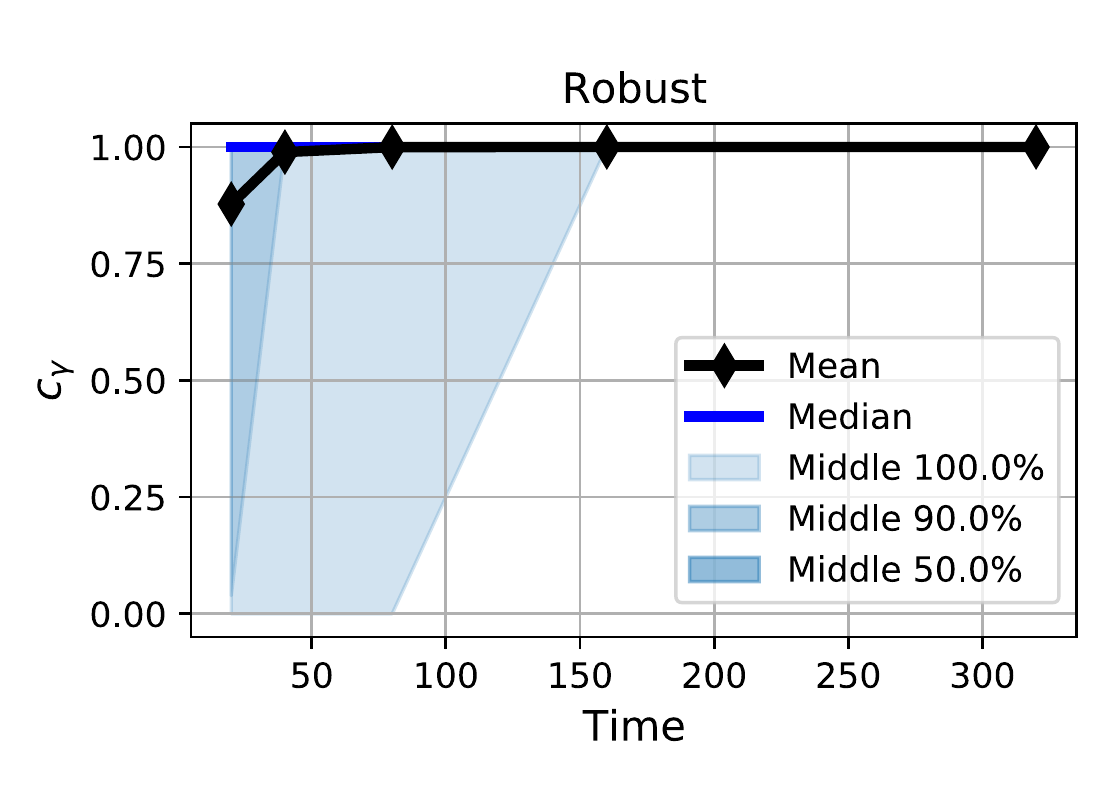}
    \caption{Scaling of multiplicative noise scale parameter $\gamma$ vs time for the example system using RMN.}
    \label{fig:gamma_matrix_scale}
\end{figure}

From Figure \ref{fig:gamma_matrix_scale} we see that at the very beginning when the uncertainty is extremely high, the multiplicative noise variance sometimes had to be reduced significantly in order to admit a solution to the generalized Riccati equation. Over time as the uncertainty decreased, the multiplicative noises were used with their native scaling almost all of the time.

\noindent
Code which realizes the algorithms of this paper and generates the reported results is available from \\
{\small
\url{https://github.com/TSummersLab/robust-adaptive-control-multinoise-output}.
}

\section{Conclusions}
We proposed a data-driven robust control algorithm that uses the bootstrap to estimate model estimate covariances and a non-conventional multiplicative noise LQG robust output feedback compensator synthesis to explicitly account for model uncertainty. 
Future work will go towards providing finite-time theoretical performance guarantees using tools from high-dimensional statistics and exploring alternative bootstrap uncertainty quantification schemes and robust control synthesis frameworks based e.g. on linear matrix inequalities and System Level Synthesis.

\acks{This material is based on work supported by the United States Air Force Office of Scientific Research under award number FA2386-19-1-4073 and the National Science Foundation under award number ECCS-2047040.}

\clearpage

\clearpage
\bibliography{refs}

\end{document}